\documentclass[prb,twocolumn,showpacs]{revtex4}
\usepackage[dvips]{graphicx}
\usepackage{latexsym}
\usepackage{amsmath}
\usepackage{amsfonts}
\usepackage{amssymb}
\usepackage{bm}
\usepackage{color}
\begin{document}
\newcommand{\fig}[2]{\includegraphics[width=#1]{#2}}
\newcommand{{\vhf}}{\chi^\text{v}_f}
\newcommand{{\vhd}}{\chi^\text{v}_d}
\newcommand{{\vpd}}{\Delta^\text{v}_d}
\newcommand{{\ved}}{\epsilon^\text{v}_d}
\newcommand{{\vved}}{\varepsilon^\text{v}_d}
\newcommand{{\bk}}{{\bf k}}
\newcommand{{\bq}}{{\bf q}}
\newcommand{{\tr}}{{\rm tr}}
\newcommand{\pprl}{Phys. Rev. Lett. \ }
\newcommand{\pprb}{Phys. Rev. {B}}

\title{Textured electronic states of the triangular lattice Hubbard model and Na$_x$CoO$_2$}
\author{Kun Jiang,$^1$ Sen Zhou,$^{2}$ and Ziqiang Wang$^1$}
\affiliation{$^1$ Department of Physics, Boston College, Chestnut Hill, MA 02467,
USA}
\affiliation{$^2$ State Key Laboratory of  Theoretical Physics, Institute of
Theoretical Physics, Chinese Academy of Sciences, Beijing 100190, China}
\date{\today}

\begin{abstract}

We show that geometric frustration and strong correlation in the triangular lattice Hubbard model lead a rich and novel phase structure of $\sqrt{3}\times\sqrt{3}$ spin-charge textured electronic states over a wide region of electron doping $0\le x \le 0.40$. In addition to the 120$^\circ$ N\'eel ordered insulator at half-filling, we found a novel spin-charge ordered insulator at $x=1/3$ with collinear antiferromagnetic (AF) order on the underlying unfrustrated honeycomb lattice. Separating the two insulating phases is a Lifshitz transition between a noncollinear AF ordered metal and one with coexisting charge order. We obtain the phase diagram and the evolution of the Fermi surface (FS). Remarkably, the correlated ground states near $x=1/3$ emerges as doping the ``1/3 AF insulator'' by excess carriers, leading to electron and hole FS pockets with important implications for the cobaltate superconducting state.


\typeout{polish abstract}
\end{abstract}


\maketitle

\section{INTRODUCTION}
Correlated electron materials with geometrically frustrated lattice structures hold great promise for novel quantum electronic states. In addition to the quantum spin liquid \cite{anderson73-87,palee08,balents10} observed in $\kappa$-organics \cite{kanoda03,kanoda05,kanoda08,matsuda08} near the Mott transition at half-filling, the sodium cobaltates Na$_x$CoO$_2$ exhibit rich and unconventional phases \cite{takada03,yywang03,schaak03,foo04,alloul09} in a wide range of electron doping $x$. Central to the properties of the sodium cobaltates is the unconventional superconducting (SC) state observed near $x=1/3$ upon water intercalation \cite{takada03}. Despite the intensive search for its possible electronic origin \cite{shastry03,baskaran03,ogata03,qhwang04,kuroki04,mazin04,motrunich04,mochizuki05,eremin08,
zhouwang08,kiesel13,ziyangmeng}, the nature and pairing mechanism of the SC phase have been a controversial and unresolved issue. Contrary to conventional wisdom, several experiments suggest that the many-electron ground state at superconducting concentrations may be in close proximity to certain hidden electronic ordered phases \cite{hasan06,yang05,balicas06,rivadulla06,matano08,cava09}. Although various ordered states have been conjectured near x=1/3 \cite{baskaran03xxx,kuroki04,motrunich04,foussats06,an06,hasan06,wrobel07,ohkawa10} and argued to be relevant for superconductivity, almost all were based on the idea of Coulomb jamming where a strong extended interaction $V$ drives a Wigner crystal-like charge ordered insulating state with a large gap to single-particle excitations which is inconsistent with experiments. The nature and the microscopic origin of the textured electronic states if they exist, and the idea that electronic fluctuation mediated superconductivity arises in their proximity have remained enigmatic due to the lack of concrete understanding of the strong correlation effect and its interplay with geometric frustration in layered triangular lattice Mott-Hubbard systems.
Even for the simplest Hubbard model, its possible electronic ground states as a function of doping have not been understood on the triangular lattice.

In this paper, we study the ground state properties and the phase diagram of the triangular lattice Hubbard model. We show that, upon electron doping, new stable phases with textured charge and spin order (both collinear and coplanar) arise as a result of geometric frustration and strong correlation and provide insights to the cobaltate unconventional normal and SC states. Specifically, we construct a spin-rotation invariant slave boson theory capable of describing both charge and noncollinear magnetic superstructures to study the ground states as a function of Hubbard $U$ and electron doping $x$. We find that adding electrons turns the frustrated 120$^\circ$ N\'eel ordered insulator at half-filling into a 3-sublattice noncollinear antiferromagnetic (AF) metal which is stable at low-doping but undergoes a Lifshitz transition accompanied by incipient charge ordering. The magnetic frustration begins to alleviate in the presence of charge inhomogeneity, and a novel AF insulator emerges at $x=1/3$ where the unfrustrated collinear spin-density resides on the underlying honeycomb lattice sites
and coexists with moderate $\sqrt{3}\times\sqrt{3}$ charge density order. We obtain the phase diagram in the regime $0\le x\le0.45$, discuss the nature of the phases and the phase transitions, and illustrate the evolution of the Fermi surface (FS).
%
Remarkably, the strongly correlated ground states near $x=1/3$ can be viewed as doping into the ``1/3 AF insulator'', giving rise to metallic phases with small electron or hole FS pockets accommodating the excess carriers. We compare our findings to recent experiments and argue that the enhanced spin and charge fluctuations together with the narrowed quasiparticle band and the nested FS pockets may have important implications for the electronic origin of the SC phase in sodium cobaltates.
\section{SPIN ROTATIONAL INVARIANT SLAVE BOSON THEORY FOR NONCOLLINEAR SPIN AND CHARGE TEXTURED STATES}
The triangular lattice Hubbard model is given by,
\begin{equation}
H=-\sum_{ij,\sigma}t_{ij}c^{\dagger}_{i\sigma} c_{j\sigma}
+U\sum_{i}{\hat n}_{i\uparrow}{\hat n}_{i\downarrow}-\mu\sum_{i\sigma} c_{i\sigma}^\dagger c_{i\sigma},
\label{hrs}
\end{equation}
where $c^{\dagger}_{i\sigma}$ creates a spin-$\sigma$ electron; $U$ is the on-site Coulomb repulsion; and ${\hat n}_{i\sigma}$ the density operator. The first three nearest neighbor hoppings $t_{ij}=(t_1,t_2,t_3)=(-202, 35, 29)$ meV produce a tight-binding dispersion with a bandwidth $W=1.34$eV for the $a_{1g}$-band in the cobaltates \cite{zhou05,zhouwang07}. To study the interplay between strong correlation and magnetic frustration, we use the Kotliar-Ruckenstein slave-boson formulation \cite{kr} with full spin-rotation invariance \cite{cli,fresard}. This strong-coupling theory correctly describes the weakly interacting limit ($U\to0$), recovers and extends the Gutzwiller approximation to the spin-rotation invariant case for all $U$ \cite{kr,cli,fresard}. By studying the spatially unrestricted solutions, we can probe inhomogeneous, textured electronic states induced by strong correlation and geometrical frustration \cite{zhouwang07}.

The local Hilbert space of the Hubbard model is represented by a spin-1/2 fermion $f_\sigma$ and six bosons: $e$ (holon), $d$ (doublon), and $p_\mu$ ($\mu=0,1,2,3$) such that an empty site $\vert0\rangle=e^\dagger \vert\text{vac}\rangle$, a doubly occupied site $|\!\!\uparrow\downarrow\rangle=d^\dagger f_\downarrow^\dagger f_\uparrow^\dagger \vert\text{vac}\rangle$, and a singly occupied site $\vert \sigma\rangle= f_{\sigma^\prime}^\dagger p_{\sigma^\prime\sigma}^\dagger\vert \text{vac}\rangle$ with sums over repeated spin indices. The spin-rotation invariance is achieved in the SU(2) representation of the $2\times2$ matrix ${\bf p}$, i.e. $p_{\sigma\sigma^\prime}^\dagger={1\over\sqrt{2}} p_\mu^\dagger \tau_{\sigma\sigma^\prime}^\mu$ where ${\bf \tau}^{1,2,3}$ and ${\bf \tau}^0$ are the Pauli and identity matrices \cite{cli}. The local spin operator ${\bf S}_i^{x,y,z}={1\over2}\tr(\tau^{1,2,3}{\bf p}_i^\dagger {\bf p_i})$. For the completeness of the Hilbert space,
\begin{equation}
Q_i=e_i^\dagger e_i+d_i^\dagger d_i+\tr({\bf p}_i^\dagger {\bf p}_i)=1.
\label{constraint1}
\end{equation}
The equivalence between the fermion and boson representations of the particle and spin density further requires,
\begin{equation}
L_i^\mu=\tr(\tau^\mu {\bf p}_i^\dagger {\bf p}_i ) + 2\delta _{\mu ,0} d_i^\dagger  d_i -\sum\limits_{\sigma \sigma '} f_{i\sigma}^\dagger  \tau_{\sigma \sigma '}^\mu f_{i\sigma '}=0.
 \label{constraint2}
\end{equation}

For electron doping, it is convenient to work in the hole-picture. Accordingly, the sign of the hopping term in Eq.~(\ref{hrs}) is reversed. Moreover, at electron doping concentration $x$, the average density of holes is given by $n=(1/N)\sum_{i\sigma} \langle f_{i\sigma}^\dagger f_{i\sigma} \rangle=1-x$. The Hamiltonian can thus be written as,
\begin{eqnarray}
H&=&\sum_{ij}
t_{ij} f^{\dagger}_{i\sigma_1} Z_{i\sigma_1\sigma}^\dagger Z_{j\sigma\sigma_2} f_{j\sigma_2}
+U\sum_{i}d_i^\dagger d_i \nonumber \\
&-&\mu\sum_{i} f_{i\sigma}^\dagger f_{i\sigma}-\sum_i\lambda_i(Q_i-1)-\sum_{i\mu}\lambda_i^\mu L_i^\mu,
\label{hsb}
\end{eqnarray}
where $\lambda_i$ and $\lambda_i^\mu$ are Lagrange multipliers enforcing the constraints
in Eqs.~(\ref{constraint1}) and (\ref{constraint2}). The renormalization factors for the hopping term has the matrix form \cite{cli,fresard}
\begin{equation}
{\bf Z}_i={\bf L}_i^{-1/2}(e_i^\dagger {\bf p}_i+{\bf {\overline p}}_i^\dagger d_i){\bf R}_i^{-1/2},
\label{zfactor}
\end{equation}
where ${\bf L_i}=(1-d_i^\dagger d_i)\tau_0-{\bf p_i}^\dagger {\bf p}_i$, ${\bf R}_i=(1-e_i^\dagger e_i)\tau_0-{\bf {\overline p_i}}^\dagger {\bf{\overline p}}_i$, and ${\bf{\overline p}}_i={\hat{\bf T}}{\bf p}_i{\hat{\bf T}}^{-1}$ is the time-reversal transformed ${\bf p}_i$.
%
%
%
The saddle-point solution of the functional-integral for Eq.~(\ref{hsb}) corresponds to condensing all boson fields $(e_i,d_i,p_{i\mu},\lambda_i,\lambda_i^\mu)$
and determining their values self-consistently by minimizing the ground state energy $\langle H \rangle$ \cite{kr,cli}.

Real space unrestricted searches for the lowest energy states indicate that, in the doping regime $0\le x\le 0.40$, the uniform paramagnetic (PM) ground state becomes unstable above a critical $U$ toward textured electronic states that always emerge with $\sqrt{3}\times\sqrt{3}$ superstructures. To determine the ground state properties of these textured phases, the phase structure and the phase transitions, it turns out to be necessary to go beyond real space calculations and study much larger systems. To this end,
we construct a superlattice formulation of the theory where each supercell contains 3 sites labeled by $\ell=1(A),2(B),3(C)$. The Hamiltonian (\ref{hsb}) becomes
\begin{eqnarray}
&H&=\sum_{\ell\ell^\prime,k}
K_{\ell\ell^\prime}(k)
 f^{\dagger}_{\ell k \sigma_1} Z_{\ell\sigma_1\sigma}^\dagger Z_{\ell^\prime\sigma\sigma_2}  f_{\ell^\prime k \sigma_2}
+U\sum_{\ell}d_\ell^\dagger d_\ell\nonumber \\
&-&\mu\sum_{\ell k} f_{\ell k\sigma}^\dagger f_{\ell k\sigma}-\sum_\ell\lambda_\ell(Q_\ell-1)-\sum_{\ell\mu}\lambda_\ell^\mu L_\ell^\mu,
\label{hsb-spercell}
\end{eqnarray}
where $k$, defined in the reduced zone, is the crystal momentum associated with the superlattice translation symmetry. The hopping matrix elements are $K_{\ell\ell^\prime}^*(k)=K_{\ell^\prime\ell}(k)$,
\begin{eqnarray}
 K_{11}&=& 2t_2
 [\cos k_+  + \cos k_- +\cos (\sqrt{3} k_x )] - \mu,
 \nonumber \\
 K_{12}&=& t_1 (1 + e^{-ik_+}  + e^{ik_-})+t_3 [2\cos(\sqrt{3}k_x ) + e^{-i3k_y }],
 \nonumber \\
 K_{13}&=& t_1 [1 + e^{-ik_+}  + e^{- i\sqrt{3} k_x}]
 \nonumber \\
 && \qquad + t_3 [2\cos k_- + e^{-i(\frac{3\sqrt{3}}{2}k_x + \frac{3}{2}k_y )}],
 \nonumber \\
 K_{23}&=& t_1 (1 + e^{-ik_-}  + e^{-i\sqrt{3} k_x })
 \nonumber \\
 && \qquad + t_3 [2\cos k_+ + e^{-i(\frac{3\sqrt{3}}{2}k_x  - \frac{3}{2}k_y )}],
 \nonumber
\end{eqnarray}
and $K_{11} = K_{22}  = K_{33}$, with $k_{\pm}={\sqrt{3}\over2}k_x\pm{3\over2}k_y$. 
Minimizing the energy leads to the following self-consistency equations at each of the 3 sites in the supercell,
\begin{eqnarray}
\sum\limits_{\ell_1,\ell_2}\frac{{\partial T_{\ell_1,\ell_2} }}{{\partial e_\ell }} &+& 2\lambda_\ell e_\ell  = 0,
\nonumber \\
\sum\limits_{\ell_1,\ell_2}\frac{{\partial T_{\ell_1,\ell_2} }}{{\partial d_\ell }} &+& (2\lambda_\ell  - 4\lambda_\ell^0 + 2U)d_\ell  = 0,
\nonumber
\\
\sum\limits_{\ell_1,\ell_2}\frac{{\partial T_{\ell_1,\ell_2} }}{{\partial p_\ell^0 }} &+& 2\lambda_\ell p_\ell^0 - 2\sum_\mu\lambda_\ell^\mu p_\ell^\mu = 0,
\nonumber \\
\sum\limits_{\ell_1,\ell_2}\frac{{\partial T_{\ell_1,\ell_2} }}{{\partial p_\ell^\alpha}}
&+&2\lambda_\ell p_\ell^\alpha - 2\lambda_\ell^0 p_\ell^\alpha-2\lambda_\ell^\alpha p_\ell^0=0, \ \alpha=1,2,3,
\nonumber
\end{eqnarray}
where $T_{\ell_1,\ell_2}=\sum_k K_{\ell_1\ell_2}(k)Z^\dagger_{\ell_1\sigma_1\sigma} Z_{\ell_2\sigma \sigma_2} \langle f_{\ell_1 k \sigma_1}^\dagger  f_{\ell_2 k\sigma_2}\rangle$ is the quantum averaged kinetic energy between sites $\ell_1$ and $\ell_2$. These equations, together with the quantum averaged constraints (\ref{constraint1}) and (\ref{constraint2}), are solved numerically
by discretizing the reduced zone with typically $600\times600$ points to allow accurate determinations of the ground state properties. We verified that all of our results are reproducible in the even larger $3\times3$ supercell calculations.

\begin{figure}
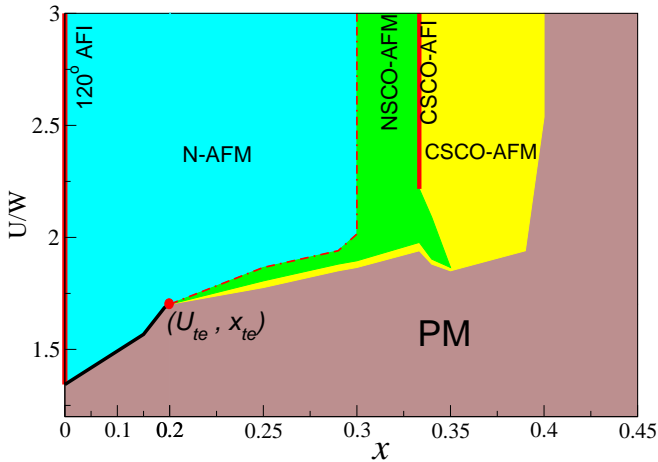

    \begin{center}
\fig{3.4in}{fig1.eps}\caption{Phase diagram of the triangular lattice Hubbard model. $(U_{\rm te},x_{\rm te})$ denotes the tetra-critical point (red circle) where the first order transition line (black), two second order transition lines, and the Lifshitz transition (dashed line) meet. Note the different horizontal scale for $x\le0.2$.}
\end{center}
\vskip-0.5cm
\end{figure}
\section{RESULTS AND DISCUSSIONS}
The obtained phase diagram is shown in Fig.~1. The stable phases in the wide region of doping $0\le x\le0.4$ are spin-charge textured electronic states for large enough $U$.
The strongly correlated electronic states are highlighted by two dramatically different insulating states at $x=0$ and $x=1/3$ (marked by red-lines). The insulating state at half-filling sets in above $U_{\rm 120}= 1.34W$ with noncollinear, 3-sublattice, $120^\circ$ N\'eel order due to magnetic frustration as shown in Fig.~2a, in good agreement with numerical renormalization group calculations \cite{yoshiyoka}. Due to the quenching of charge fluctuations at large $U$ at half-filling, the charge density is uniform.
Remarkably, at $x=1/3$, a novel textured insulating state emerges above $U_{c2}=2.22W$ with {\em unfrustrated} collinear AF order on the underlying honeycomb lattice as shown in Fig.~2b. The avoided magnetic frustration in this ``1/3 AF insulator'' is achieved via moderate $\sqrt{3}\times\sqrt{3}$ charge order: on one of the 3 sublattices, the charge density is larger and the spin density vanishes. We first describe the evolution of ground states between these strong coupling insulators as a function of $x$, and then study the transitions in the ground state at a fixed doping as a function of $U$.

\begin{figure}
    \begin{center}
\fig{3.2in}{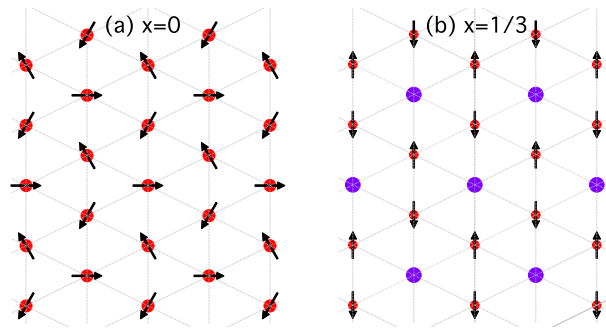}\caption{Magnetic ordered insulating states at large U. (a) $120^\circ$ noncollinear N\'eel order at $x=0$ and $U=2W$. (b) Unfrustrated AF order on the underlying honeycomb lattice with charge order at $x=1/3$ and $U=3W$. Solid circles indicate the charge density.}
\end{center}
\vskip-0.5cm
\end{figure}
%
%

It is instructive to start with the 3-sublattice 120$^\circ$ AF insulator (120$^\circ$-AFI). It originates from the geometrically frustrated AF correlation on the triangular lattice. The noncollinear magnetic order splits the 3 subbands into 6 spin-nondegenerate bands with the lowest three filled in the half-filled insulating state. Electron doping leads to the occupation of the fourth band, and the noncollinear AF metal (N-AFM) emerges with an electron FS enclosing the zone center ($\Gamma$ point). With increasing $x$, the FS grows with a volume of $x$ and the ordered moments decrease due to carrier hopping. The subband gaps are reduced accordingly but are nonzero and the N-AFM remains stable for a wide doping range as seen in Fig.~1 until the growing hexagonal FS begins to make point-contact with the $\sqrt{3}\times\sqrt{3}$ reduced zone boundary form the inside near $x\simeq0.3$ and a Lifshitz transition takes place through umklapp scattering (dotted-dash line in Fig.~1). Fig.~3a and 3b display the FS before and after the transition, showing the FS topology change and the emergence of small hole FS pockets across the Lifshitz transition. It should be noted that although there is no additional lattice symmetry breaking associated with the Lifshitz transition, the $\sqrt{3}\times\sqrt{3}$ charge order becomes prominent as do the deviations of the spin-density on the 3-sublattices from the 120$^\circ$ order, when the system enters the noncollinear spin-charge ordered AF metal (NSCO-AFM) phase shown in Fig.~1. Interestingly, the emergence of charge inhomogeneity allows the alleviation of magnetic frustration in the NSCO-AFM phase and the collinear spin-charge ordered AF metal (CSCO-AFM) with AF order on the unfrustrated honeycomb lattice eventually prevails for $x>1/3$. At $x=1/3$, the lower two of the three spin-degenerate bands are filled with 4 electrons per unit cell, leading to the ``1/3 AF insulator'', which we denote as collinear spin-charge ordered AF insulator (CSCO-AFI).
\begin{figure}
    \begin{center}
\fig{3.2in}{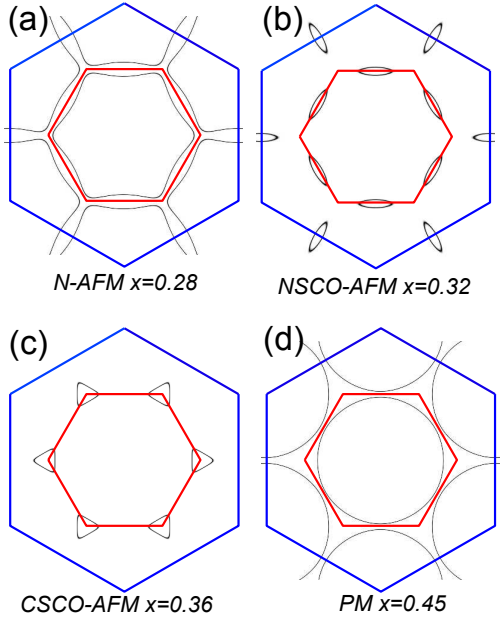}\caption{FS of electronic textured phases at U=3W and doping (a) $x=0.28$. (b) $x=0.32$. (c) $x=0.36$ (d)$x=0.45$.
}
\end{center}
\vskip-0.5cm
\end{figure}

Next, we turn to the phase transitions as a function of $U$ at a fixed doping. At half-filling, a first order transition separates the PM metal from the 120$^\circ$-AFI with a two-component magnetic order parameter. We find that the first order line extends and terminates at a tetra-critical point $(U_{\rm te},x_{\rm te})=(1.7W,0.2)$. For $x>x_{\rm te}$, the first order line splits into three continuous transitions with increasing $U$ as shown in Fig.~1: PM $\rightarrow$ CSCO-AFM $\rightarrow$ NSCO-AFM $\rightarrow$ N-AFM. The origin of the tetra-critical point has to do with the FS of the {\em PM metal} making contact with the reduced zone boundary from the outside at $x_{\rm te}$. The latter induces $\sqrt{3}\times\sqrt{3}$ charge order through umklapp scattering, which enables the magnetic order parameters to develop successively in the CSCO-AFM and the NSCO-AFM phases. Increasing $U$ further for $0.2<x<0.3$, the NSCO-AFM phase meets the phase boundary of the Lifshitz transition to the N-AFM phase as the FS pockets overlap and transform into the hole FS centered around $\Gamma$-point shown in Fig.~3a.
%

\begin{figure}
    \begin{center}
\fig{3.4in}{fig4.eps}\caption{(a) Schematic phase diagram at $x=1/3$ as a function of $U$. The evolution of the charge density, magnitude and orientation of the spin density on the 3 sublattices sketched in (a) are shown quantitatively in (b), (c), and (d) respectively.}
\end{center}
\vskip-0.5cm
\end{figure}
\begin{figure}
    \begin{center}
\fig{3.4in}{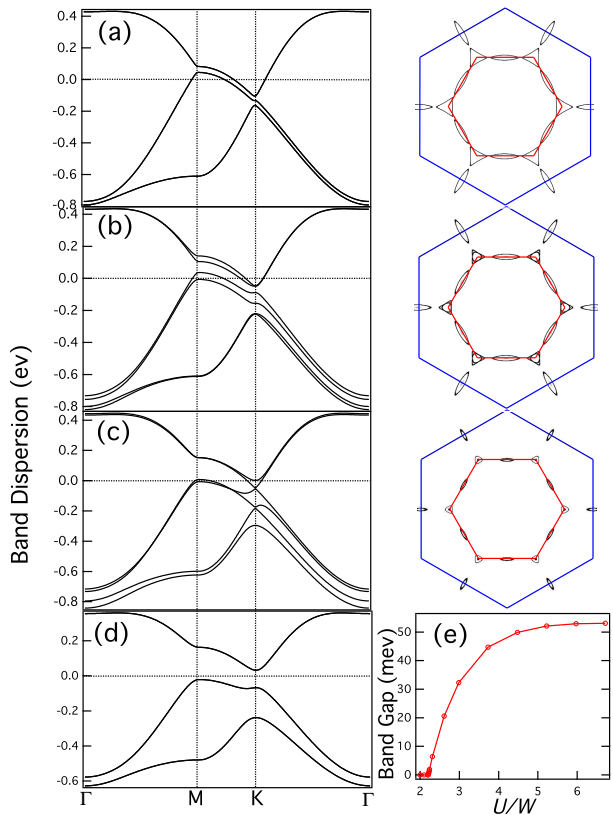}\caption{Band dispersion (left panel) and FS topology at $x=1/3$:(a) CSCO-AFM at U=1.97W, (b) NSCO-AFM at U=2.08W, (c) C-FRM at U=2.18W, and (d) CSCO-AFI at U=6.7W. The single-particle gap in the CSCO-AFI phase is shown in (e) as a function of $U/W$.}
\end{center}
\vskip-0.5cm
\end{figure}
In Figs. 4 and 5, we provide quantitative results on the phase evolution at $x=1/3$. With increasing $U$, the PM metal becomes unstable and makes a transition at $U_{c1}=1.94W$ into the CSCO-AFM phase, where gaps open due to umklapp scattering along the $M-K$ and the $K-\Gamma$ directions as shown in Fig.~5a, producing three subbands in the folded zone and truncating the FS into six electron and hole pockets. The electronic texture (Fig.~4a) is identical to the one in the CSCO-AFI phase above $U_{c2}$.
As shown in Figs.~4b-d, sublattice A has a higher charge density but zero spin density, whereas collinear AF ordered spin moments reside on sublattices B and C with lower charge densities, forming an underlying honeycomb lattice. One would have expected that this charge-spin ordered semimetal (SM) phase to evolve continuously into the CSCO-AFI as the magnitude of the order parameters increases with increasing $U$, thus gapping out the entire FS. However, this is not the case. This SM phase is stable only in a small region (see Fig.~1) until $U_{cp}=1.98W$ above which noncollinear (coplanar), two-component magnetic order emerges; a magnetic moment develops on sublattice A while the existing moments on sublattices B and C cant away from $180^\circ$ (Fig.~4a). Due to the noncollinearity of the magnetic order, the 3 spin-degenerate bands split into six shown in Fig.~5b in this NSCO-AFM phase. The evolution of the charge and spin density on the 3 sublattices, $n_\ell$ and $m_\ell$,  as well as the relative angles between the ordered spin moments $\theta_{\ell\ell^\prime}$ are shown in Figs.~4b-d as a function of $U$. This NSCO-AFM phase spans a wider region $1.98W <U <2.15W$. Due to the interplay of the charge and spin degrees of freedom, $n_\ell$, $m_\ell$, and $\theta_{\ell\ell^\prime}$ are nonmonotonic functions and show intricate evolutions with $U$. With the emergence of $m_A$, the noncollinear magnetic order first moves towards the $120^\circ$ state ($\theta_{\ell\ell^\prime} \to 120^\circ$), but quickly reverses path since the growing $m_{B,C}$ accompanying the decrease of $n_{B,C}$ prefers to be AF correlated ($\theta_{BC}\to 180^\circ$) while $\theta_{AB}$ remains degenerate with $\theta_{AC}$. In order to alleviate frustration, the charge density $n_A$ continues to increase such that $m_A$ reduces. As can be seen in Fig.~4b-d, surprisingly, the path toward the CSCO-AFI above $U_{c2}$ is interrupted by an incipient collinear ferrimagnetic metal (C-FRM) phase at $U_{FR}=2.15W$, where $n_C$$(n_B)$ increases (decreases) sharply such that $n_C\simeq n_A > n_B$ and $m_C\simeq m_A < m_B/2$. To minimize frustration, the larger spin moment $m_B$ is AF correlated with the smaller and parallel $m_A$ and $m_C$ ($\theta_{AB}=\theta_{BC}=180^\circ$, $\theta_{AC}=0$). The net ferromagnetic moment splits the spin degeneracy such that there remains six quasiparticle bands shown in Fig.~5c. The C-FRM phase is stable until $U_{c2}$ where a redistribution of the charge/spin density takes place to further minimize magnetic frustration: $n_A$ increases to 1.36 and $m_A$ decreases to zero; while $n_B$ and $n_c$ approaches the common value of $1.32$ and $m_B$ and $m_C$ to $0.18$ in the large U limit.
An insulating gap opens as the system enters the CSCO-AFI phase as shown in Fig.~5d-e, which is the stable phase for $U > U_{c2}$.  Compared to the CSCO-AFM phase just above $U_{c1}$, the spin moments on $B$ and $C$ sublattices have grown and rotated by $90^\circ$ above $U_{c2}$.
We stress that the charge ordering necessary for the emergence of these textured states near $x=1/3$ arises from the Lifshitz transition
and is very different from the $\sqrt{3}\times\sqrt{3}$ Wigner crystal-like state driven by Coulomb jamming due to a large next-nearest neighbor $V$ \cite{motrunich04}. Moreover, the ``1/3 AF insulator'' is different from the fully charge-disproportionate state with a large insulating gap proposed in LSDA+U calculations \cite{pickett}. Indeed, as shown in Fig.~5e, the small excitation gap in the CSCO-AFI phase opens at $U_{c2}$ and only reaches about $53$meV in the large-$U$ limit.

It is remarkably that the spin-charge textured ground states occupy such a significant portion of the phase diagram around $x=1/3$. Indeed, the large-$U$ phase structure can be generically understood as either electron ($x>1/3$) or hole ($x<1/3$) doping into the corresponding ``1/3 AF insulator'', leading to correlated metallic phase with nested electron or hole FS pockets in Figs.~3(b) and 3(c). For example, for $x>1/3$, the excess carriers give rise to the CSCO-AFM metal phase with electron FS pockets centered around the zone corners. As shown in Figs.~ 3(c) and 3(d), the latter grow with increasing $x$ until they touch and coalesce to trigger a transition into the uniform PM phase above $x=0.4$ in the phase diagram Fig.~1.
%
\section{CONCLUSIONS}
We conclude with a discussion of the implications on the sodium cobaltates. Theoretical estimates \cite{zhou05,ishida05,gtwang08,shorikov11} and the valence band resonant photoemission \cite{hasan04} suggest $U=3\sim5$eV for the Co $d$-electrons typical of $3d$ transition metals. Together with the bandwidth $W\simeq1.34$eV, the value of $U/W=2.2 - 3.7$ puts the cobaltates near $x=1/3$ in the regime of the textured states on the phase diagram with small electron and/or hole FS pockets. There are experimental indications from ARPES that the PM phase with the large $a_{1g}$ FS is in proximity to such hidden ordered phases \cite{hasan06,yang05}. Moreover, quantum oscillations find remarkably small FS pockets at $x=0.3$
possibly due to electronic superstructures \cite{balicas06}. The main reason that such states have not been widely observed in unhydrated cobaltates is likely due to the disordered Na dopant ions \cite{noteonu}. Indeed, magnetic susceptibility measurements in thermally annealed samples around $x=0.36$ find evidence for a magnetic ordered state \cite{rivadulla06}. We believe that water intercalation expands the c-axis and provides
effective screening of the dopant potential, making the physical properties more suitable for the 2D triangular lattice Hubbard model description.
Indeed, NMR experiments find that the principal effect of hydration is to reveal enhanced spin fluctuations at low temperatures, when compared to unhydrated single crystals at the same nominal Na concentrations \cite{matano08}. More direct evidence supporting this view comes from hydrated samples at $x\simeq0.3$, where a specific heat anomaly observed at a critical temperature near $7$K was unaffected by a $9$T magnetic field
and identified as associated with density wave order \cite{cava09}.
We thus propose that the cobaltates near $x=1/3$ are in proximity to such ``hidden'' textured phases with spin and charge order and the enhanced electronic fluctuations can mediate the SC pairing interaction.

\section{ACKNOWLEDGMENTS}
This work is supported in part by DOE DE-FG02-99ER45747 and NSF DMR-0704545. ZW thanks Aspen Center for Physics for hospitality.

%



\begin{thebibliography}{99}
\bibitem{noteonu}
We take the point of view that the vlaue of $U$ does not change significantly with the doping level and the usual properties in the sodium rich phases originate from the interplay of strong correlation and sodium dopant order.
\end{thebibliography}
\end{document}